\documentclass[conference]{IEEEtran}
\IEEEoverridecommandlockouts
\usepackage{cite}
\usepackage{amsmath,amssymb,amsfonts}
\usepackage{algorithmic}
\usepackage{graphicx}
\usepackage{textcomp}
\usepackage{xcolor}

\usepackage{algorithm}
\usepackage{enumitem}
\usepackage{diagbox}     
\usepackage{makecell}    
\usepackage{tabularx}
\usepackage{booktabs}   
\usepackage{multirow}

\usepackage{colortbl}     
\usepackage{array}         
\usepackage{arydshln}

\newcolumntype{C}[1]{>{\centering\let\newline\\\arraybackslash\hspace{0pt}}m{#1}}
\newcolumntype{G}[1]{>{\columncolor{lightgreen}[\tabcolsep][\tabcolsep]\centering\let\newline\\\arraybackslash\hspace{0pt}}m{#1}}

\definecolor{lightgreen}{HTML}{E6F7EB}
\def\BibTeX{{\rm B\kern-.05em{\sc i\kern-.025em b}\kern-.08em
    T\kern-.1667em\lower.7ex\hbox{E}\kern-.125emX}}
\begin{document}

\title{Graph-Guided Adaptive Channel Elimination for KV Cache Compression\\
% {\footnotesize \textsuperscript{*}Note: Sub-titles are not captured for https://ieeexplore.ieee.org  and
% should not be used}
\thanks{\textsuperscript{*}Corresponding author: Yuanchao Bai (Emails: yuanchao.bai@hit.edu.cn). This work was supported in part by National Natural Science Foundation of China under Grants 62301188, 92270116 and U23B2009, in part by China Postdoctoral Science Foundation under Grant 2022M710958, and in part by Heilongjiang Postdoctoral Science Foundation under Grant LBH-Z22156.}
}

\author{
    \IEEEauthorblockN{
        Enwei Tong\textsuperscript{1},
        Yao Zhu\textsuperscript{2},
        Yuanchao Bai\textsuperscript{1*},
        Kai Wang\textsuperscript{1},
        Xianming Liu\textsuperscript{1},
        and Xiangyang Ji\textsuperscript{3}
    }

    \IEEEauthorblockA{\textsuperscript{1}Faculty of Computing, Harbin Institute of Technology, Harbin, China}
    \IEEEauthorblockA{\textsuperscript{2}Chu Kochen Honors College, Zhejiang University, Hangzhou, China}
    \IEEEauthorblockA{\textsuperscript{3}Department of Automation, Tsinghua University, Beijing, China}

    % \IEEEauthorblockA{
    %     Emails: \{24S103434, cswangkai\}@stu.hit.edu.cn,
    %     \{yuanchao.bai, csxm\}@hit.edu.cn, \\
    %     ee\_zhuy@zju.edu.cn,
    %     xyji@tsinghua.edu.cn
    % }

    % \IEEEauthorblockA{\textsuperscript{*}Corresponding author: yuanchao.bai@hit.edu.cn}
}

\maketitle

\begin{abstract}
Large Language Models have revolutionized natural language processing, achieving unprecedented success across a vast range of tasks. However, their practical application in long-context scenarios is severely hampered by the formidable memory footprint of the Key-Value cache. While channel pruning has emerged as a promising compression strategy, existing methods evaluate channel importance in isolation, fundamentally ignoring the inter-channel interactions that collectively dictate model performance. This oversight leads to suboptimal pruning decisions. To address this, we introduce \textbf{GRACE} (\textbf{GR}aph-guided \textbf{A}daptive \textbf{C}hannel \textbf{E}limination), a novel framework that reframes KV cache compression as a graph-based optimization problem. GRACE models channels as nodes and their interactions as weighted edges, enabling the identification of a near-optimal channel subset for pruning by minimizing the reconstruction error of the attention weight matrix. Furthermore, GRACE incorporates an adaptive protection mechanism that shields salient key channels from removal, ensuring a robust autoregressive decoding process. Extensive experiments show that GRACE can reduce KV cache size by 60\% with negligible performance degradation, consistently outperforming the state-of-the-art method.
\end{abstract}


\section{Introduction}
\label{sec:intro}

Large Language Models (LLMs)~\cite{touvron2023llama, jiang2023mistral7b} have established a new paradigm in natural language processing, delivering state-of-the-art performance in generation~\cite{mo2024largelanguagemodelllm}, reasoning~\cite{yuan2024advancingllmreasoninggeneralists}, and contextual understanding~\cite{zhang2019hibertdocumentlevelpretraining}. Their success is largely underpinned by the scaling law principle, which demonstrates that emergent capabilities arise from increasing model size and training data~\cite{kaplan2020scaling}. In parallel with scaling parameters and data, expanding the context window size has become a critical frontier for enhancing LLMs' capabilities~\cite{team2024gemini}.

Although the scaling paradigm has unlocked remarkable capabilities, it has also introduced considerable computational and memory challenges during inference.
This overhead is primarily driven by the Key-Value (KV) cache, whose memory footprint grows linearly with the sequence length.
This linear scaling presents a prohibitive bottleneck in long-context scenarios, severely constraining the model's ability to efficiently process extended sequences.

To mitigate this memory pressure, various strategies have been explored.
Model-level approaches enable different query heads to share KV caches~\cite{shazeer2019fast, ainslie2023gqa, brandon2024reducing}.
Token-level methods selectively evict entries from the cache~\cite{zhang2023h2o, li2024snapkv, xiao2024efficient, cai2024pyramidkv}, while sparse retrieval techniques retain the full KV cache but selectively read relevant entries during inference~\cite{tang2024quest, chen2024magicpig, liu2024retrievalattention}.
Finally, quantization reduces memory usage by lowering the precision of the KV cache~\cite{xiao2023smoothquant, hooper2024kvquant, liu2024kivi, zhang2024q}.

\begin{figure}[!t]
  \centering
  \centerline{\includegraphics[width=\linewidth]{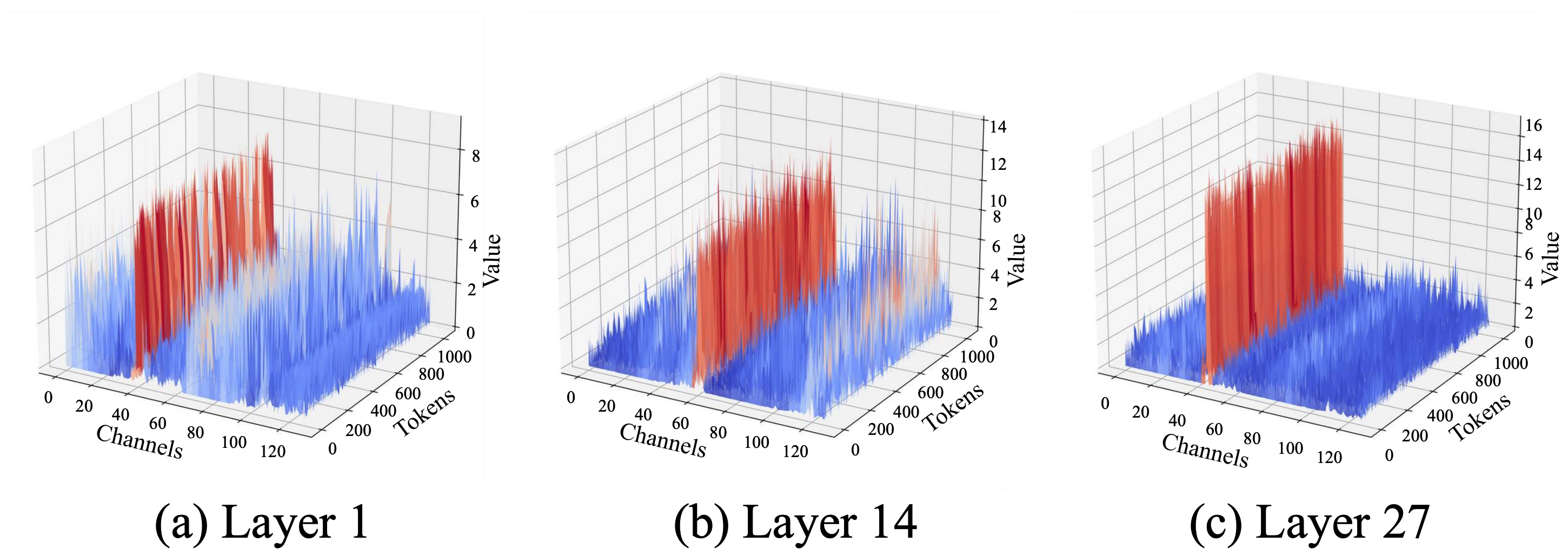}}
\caption{Magnitude visualization of key cache. A small subset of channels concentrates the majority of the energy.}
\label{fig:key_3d_visual}
\end{figure}

Orthogonal to these approaches, channel-level pruning has emerged as a promising technique.
Existing methods, however, suffer from fundamental limitations. THINK~\cite{xu2024think} evaluates the importance of each channel independently, neglecting the inter-channel interactions that jointly determine attention reconstruction quality. LeanK~\cite{zhang2025leank}, on the other hand, relies on offline training to learn static pruning masks, which limits its practicality. Furthermore, neither method considers the discrepancy between the query vectors used during inference and those from the prefilling stage, further degrading performance.
% Orthogonal to these approaches, channel-level pruning~\cite{xu2024think, zhang2025leank} has emerged as a promising technique.
% However, existing methods evaluate the importance of each channel in isolation, fundamentally neglecting the critical impact of inter-channel interactions on model effectiveness.
% Furthermore, they often overlook the discrepancy between the query vectors used during inference and those from the prefilling stage, further degrading performance.

In this paper, we propose \textbf{GRACE} (\textbf{GR}aph-guided \textbf{A}daptive \textbf{C}hannel \textbf{E}limination), a novel framework that reimagines channel pruning through the lens of graph theory. Unlike prior works, GRACE models channels as nodes and their interactions as edges, using a specialized greedy strategy to minimize attention reconstruction error.
Our contributions are summarized as follows:
\begin{itemize}
    \item \textbf{Graph-Theoretic Formulation:} We introduce \textbf{MIES} (Minimum Incremental Error Selection), a greedy algorithm that captures second-order channel interactions, overcoming the limitations of independent scoring.
    \item \textbf{Training-Free:} GRACE is entirely training-free and plug-and-play, allowing for flexible deployment across various pruning ratios and models.
    \item \textbf{Outlier Protection:} We design an adaptive mechanism to preserve salient channels, ensuring robustness against distribution shifts.
\end{itemize}

\section{Related Work}

\subsection{Efficient KV Cache Management}
To alleviate memory bottlenecks in long-context inference, strategies have been proposed across various granularities.

\textbf{Model-Level Optimization.} Architectural innovations like MQA, GQA and CLA reduce memory footprints by physically sharing Key-Value heads across multiple Query heads, thereby decreasing the cache size by a factor proportional to the grouping ratio~\cite{shazeer2019fast, ainslie2023gqa, brandon2024reducing}.

\textbf{Token-Level Eviction.} Token eviction policies maintain a fixed budget by discarding non-essential tokens. Methods like H2O and SnapKV greedily retain ``heavy hitters'' based on historical attention scores~\cite{zhang2023h2o, li2024snapkv}, while StreamingLLM ensures stability by preserving initial ``attention sink'' tokens~\cite{xiao2024efficient}. Additionally, PyramidKV optimizes efficiency by dynamically allocating cache sizes across layers~\cite{cai2024pyramidkv}.

\textbf{Sparse Retrieval.} Distinct from permanent eviction, sparse retrieval strategies retain the full KV cache in off-chip memory but load only relevant entries during inference~\cite{tang2024quest, chen2024magicpig, liu2024retrievalattention}.

\textbf{Quantization.} Quantization techniques compress memory usage by converting the cache into low-bit representations~\cite{xiao2023smoothquant, hooper2024kvquant, liu2024kivi, zhang2024q}. Variations include applying asymmetric strategies to Key and Value matrices~\cite{liu2024kivi} or maintaining higher precision for critical tokens~\cite{zhang2024q}.

\subsection{Channel-Level Pruning}
Recent studies indicate significant redundancy within the K cache channels, identifying that many channels contribute minimally to long-context inference~\cite{xiao2024efficient, xu2024think, hong2024token}. Consequently, channel pruning has emerged as a promising direction, which is orthogonal to the aforementioned strategies and can be combined for enhanced efficiency.

THINK~\cite{xu2024think} pioneered this field by formulating pruning as reconstruction error minimization. However, it evaluates channels independently, ignoring critical inter-channel covariance, which leads to suboptimal selection.
LeanK~\cite{zhang2025leank} proposes a framework that trains static masks to prune channels. While LeanK captures some structural information, it relies on offline training to produce fixed masks, limiting its adaptability to dynamic compression ratios or deployment changes.

Furthermore, recent research reveals that a small fraction of channels with massive magnitudes—often termed ``outliers''—exert a disproportionate impact on model outputs~\cite{sun2024massive, dettmers2022llmint88bitmatrixmultiplication}. Moreover, during the decoding process, the distribution of query vectors can shift significantly. These phenomena imply that specific K cache channels must be explicitly preserved to maintain robustness against such shifts, a critical aspect that has been largely overlooked in existing channel pruning literature.

\begin{figure}[!t]
  \centering
  \centerline{\includegraphics[width=\linewidth]{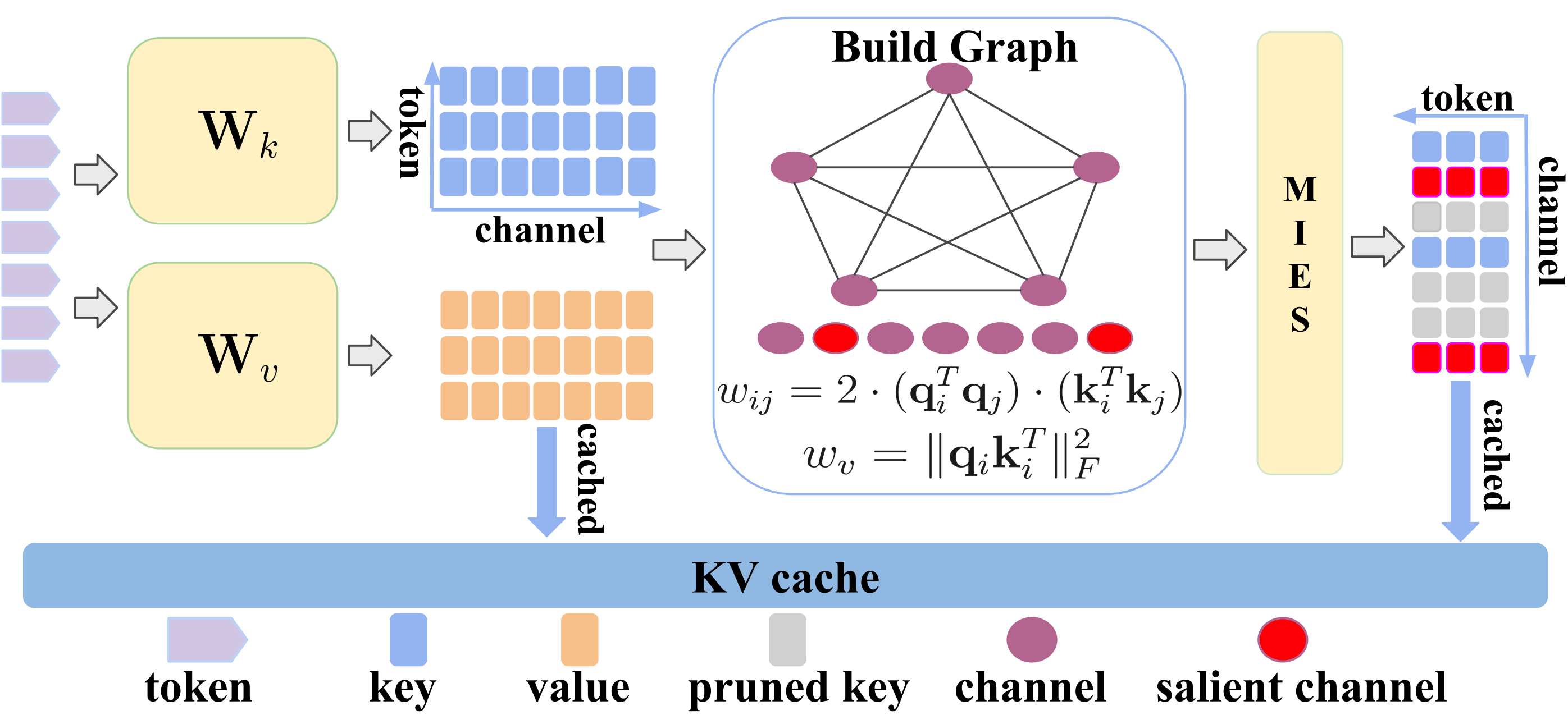}}
\caption{An overview of the proposed GRACE framework.}
\label{fig:overview}
\end{figure}

\section{The Proposed GRACE Framework}
\label{sec:method}
\subsection{Preliminary}
\label{sec:preliminary}
During the inference phase, LLMs cache previous key and value tensors to prevent redundant computations. This caching mechanism operates identically on every decoder layer and attention head. For notational simplicity, we henceforth omit the layer and head indices.

At a given decoding step $t$, the input $\mathbf{X}_{t} \in \mathbb{R}^{L\times d}$, where $L$ denotes the sequence length and $d$ is the embedding dimension, is projected by matrices $\mathbf{W}_{q},\mathbf{W}_{k}, \mathbf{W}_{v} \in \mathbb{R}^{d\times d}$ to obtain the query, key, and value tensors: $\mathbf{Q}_{t}=\mathbf{X}_{t}\mathbf{W}_{q}, \mathbf{K}_{t} = \mathbf{X}_{t}\mathbf{W}_{k}, \mathbf{V}_{t}=\mathbf{X}_{t} \mathbf{W}_{v}$. Subsequently, the key and value tensors are appended to their respective caches: $\mathcal{K}_{t}=\{\mathcal{K}_{t-1},\mathbf{K}_{t}\}, \mathcal{V}_{t}=\{\mathcal{V}_{t-1},\mathbf{V}_{t}\}$,
where $\mathcal{K}_{t}$  and $\mathcal{V}_{t}$ represent the cached key and value states at step $t$.

During the initial prefilling stage ($t=0$), the model processes a long input sequence to capture intricate semantic dependencies and establish the necessary contextual background. Consequently, the resulting initial key and value tensors, $\mathbf{K}_{0}$ and $\mathbf{V}_{0}$, constitute the vast majority of the cache. Compressing these initial tensors is therefore a pivotal strategy for alleviating memory bottlenecks and improving the inference efficiency of LLMs.

As illustrated in Fig.~\ref{fig:key_3d_visual}, a small subset of channels consistently exhibits large magnitudes, while many other channels have near-zero magnitudes across almost all tokens. Notably, this phenomenon is not as apparent in the value cache \cite{liu2024kivi}. THINK \cite{xu2024think} pioneered a channel-level approach. It prunes a subset of channels from $\mathbf{K}_{0}$  by minimizing the Frobenius norm of the difference in the attention matrix before and after pruning. This objective is formalized as follows:
\begin{equation}
\label{eq:raw_opt}
\begin{aligned}
\min_{\mathbf{S}}& \quad \left\|\mathbf{Q}_{0}^{obs}\mathbf{K}_{0}^{T}-\mathbf{Q}_{0}^{obs}\mathbf{S}\mathbf{K}_{0}^T\right\|_F\\
\text{subject to}& \quad \operatorname{Trace}(\mathbf{S})=\lfloor(1-\lambda)d\rfloor\\
                 & \quad ~\mathbf{S}=\operatorname{diag}(s_1,\ldots,s_d), \text{where}~s_j\in\{0,1\}
\end{aligned}
\end{equation}
Here, $\mathbf{Q}_{0}^{obs}$  represents the query vectors within observation window of $\mathbf{Q}_{0}$. To solve this optimization problem, THINK assigns a score to each channel $j$ via $\text{score}_j=\| \mathbf{Q}_{0}^{obs}[:,j]\mathbf{K}_{0}[:,j]^{T}\|_{F}$ and retains only the channels with the highest scores. % A key limitation of this method is its greedy nature: it assesses the score of each channel in isolation. By ignoring the inter-channel dependencies and the collective impact of pruning, the approach yields a suboptimal result.
\subsection{Motivation for a Graph-Theoretic Approach}
\begin{figure}[!t]
  \centering
  \centerline{\includegraphics[width=\linewidth]{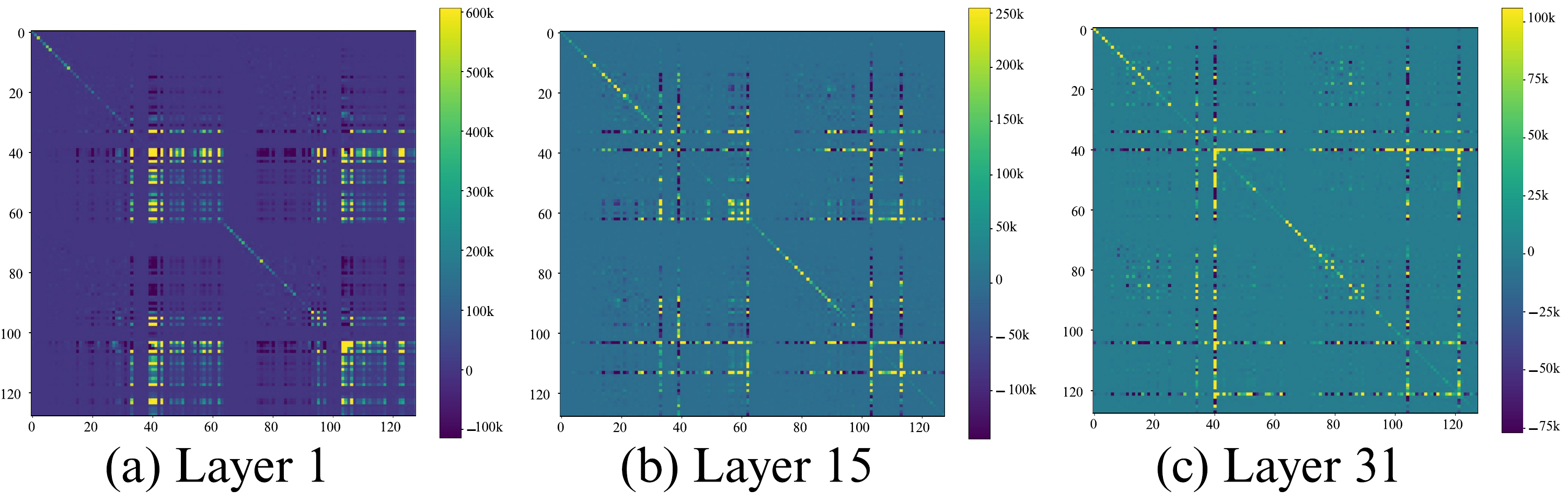}}
\caption{Heatmap of interaction terms between channels, where $\mathrm{heatmap}_{i,j} = \mathbf{k}_{i}^{T}\mathbf{k}_{j} \times \mathbf{q}_{i}^{T}\mathbf{q}_{j}$. Off-diagonal terms are non-negligible.}
\label{fig:heatmap}
\end{figure}

As illustrated in Fig.~\ref{fig:overview}, our proposed GRACE framework introduces a two-stage approach to KV cache compression. First, we identify and shield a subset of inherently salient channels to ensure robustness. A complete, weighted graph is then constructed from the remaining candidate channels to explicitly model their inter-channel interactions. Second, we employ our MIES algorithm (detailed in Algorithm \ref{alg:iap_pruning}) on the constructed graph, iteratively selecting the one that minimally increases the attention reconstruction error.

\begin{algorithm}[t]
\caption{\textbf{M}inimum \textbf{I}ncremental \textbf{E}rror \textbf{S}election}
\label{alg:iap_pruning}
\renewcommand{\algorithmicrequire}{\textbf{Input:}}
\renewcommand{\algorithmicensure}{\textbf{Output:}}
\renewcommand{\algorithmiccomment}[1]{\STATE \textit{// #1}}
\begin{algorithmic}[1]
\REQUIRE Query observation matrix $\mathbf{Q}_0^{obs} \in \mathbb{R}^{L_{obs} \times d}$, Key cache matrix $\mathbf{K}_0 \in \mathbb{R}^{L \times d}$, Pruning ratio $\lambda$.
\ENSURE $\mathcal{D}_{prune}$: Set of pruned channel indices.

\STATE Let $\mathbf{q}_j = \mathbf{Q}_0^{obs}[:, j]$ and $\mathbf{k}_j = \mathbf{K}_0[:, j]$ for $j \in \{0, \dots, d-1\}$.
\STATE $n_{prune} \gets \lceil \lambda d \rceil$
\STATE $\mathcal{D}_{cand} \gets \{0, 1, \dots, d-1\}$, $\mathcal{D}_{prune} \gets \emptyset$

\COMMENT{Initialize scores with each channel's self-importance.}
\FOR{each $j \in \mathcal{D}_{cand}$}
    \STATE $s_j \gets \| \mathbf{q}_j \mathbf{k}_j^T \|_F^2$
\ENDFOR
\COMMENT{Iteratively prune channels that minimally increase the cumulative error.}
\FOR{$i$ from $1$ to $n_{prune}$}
    \STATE $j^* \gets \arg\min_{j \in \mathcal{D}_{cand}} s_j$.
    \STATE $\mathcal{D}_{prune} \gets \mathcal{D}_{prune} \cup \{j^*\}$.
    \STATE $\mathcal{D}_{cand} \gets \mathcal{D}_{cand} \setminus \{j^*\}$.
    \COMMENT{Update remaining scores with inter-channel interactions.}
    \FOR{each $k \in \mathcal{D}_{cand}$}
        \STATE $s_k \gets s_k + 2 \cdot (\mathbf{q}_k^T \mathbf{q}_{j^*}) \cdot (\mathbf{k}_k^T \mathbf{k}_{j^*})$.
    \ENDFOR
\ENDFOR

\RETURN $\mathcal{D}_{prune}$
\end{algorithmic}
\end{algorithm}

To understand the theoretical motivation behind this graph-based approach, we first decompose the pruning objective in \eqref{eq:raw_opt} to reveal the underlying error structure. By expanding the Frobenius norm, we can gain a more detailed understanding of how pruning a subset of channels impacts the reconstruction of the attention matrix. Let $\mathcal{D}_k=\{i|s_{i}=1\}, \mathcal{D}_p=\{i|s_{i}=0\}$ represent the index sets of kept and pruned channels respectively, and let $\mathbf{q}_{i},\mathbf{k}_{i}$ denote the $i$-th columns of $\mathbf{Q}_{0}^{obs}$ and $\mathbf{K}_{0}$, respectively. Then we can obtain:
\begin{equation}
\label{eq:decomposed_loss}
\begin{aligned}
&\|\mathbf{Q}_0^{obs} \mathbf{K}_0^T - \mathbf{Q}_0^{obs} \mathbf{S} \mathbf{K}_0^T\|_F^2 = \left\|\sum_{i \in \mathcal{D}_p} \mathbf{q}_i \mathbf{k}_i^T\right\|_F^2 \\
&= \textbf{Trace}\left(\left(\sum_{i \in \mathcal{D}_p} \mathbf{q}_i \mathbf{k}_i^T\right) \left(\sum_{i \in \mathcal{D}_p} \mathbf{q}_i \mathbf{k}_i^T\right)^T\right) \\
&= \sum_{i \in \mathcal{D}_p} \sum_{j \in \mathcal{D}_p} \textbf{Trace}(\mathbf{q}_i \mathbf{k}_i^T \mathbf{k}_j \mathbf{q}_j^T) \\
&=  \sum_{i \in \mathcal{D}_{p}} \sum_{j \in \mathcal{D}_{p}} (\mathbf{q}_i^T \mathbf{q}_j) \cdot (\mathbf{k}_i^T \mathbf{k}_j) \\
&= \sum_{i \in \mathcal{D}_{p}} \| \mathbf{q}_i \mathbf{k}_i^T \|_F^2+ \sum_{i \in \mathcal{D}_{p}} \sum_{j \in \mathcal{D}_{p} \setminus \{i\} } (\mathbf{q}_i^T \mathbf{q}_j) \cdot (\mathbf{k}_i^T \mathbf{k}_j)
\end{aligned}
\end{equation}
This decomposition provides critical insight. The total pruning error is not simply the sum of losses from each individually pruned channel (the terms $i=j$), but is the sum of all terms of self-importance and cross-channel interaction (the terms $i \neq j$) within the pruned set $\mathcal{D}_{p}$. Prior methods that score channels independently, such as THINK \cite{xu2024think}, effectively only consider the diagonal of this error matrix ($\sum_{i \in \mathcal{D}_{p}} \| \mathbf{q}_i \mathbf{k}_i^T \|_F^2$), thereby ignoring the significant contributions from inter-channel interactions. Fig.~\ref{fig:heatmap} reveals that the inter-channel interactions (the off-diagonal elements) have substantial values, indicating that incorporating them into the pruning process leads to performance improvements.

\subsection{Graph-based Problem Formulation and Solution}
The presence of cross-channel interaction terms in \eqref{eq:decomposed_loss} motivates a graph-theoretic formulation of the pruning problem. We model the task as finding a minimum weight subgraph in a complete, weighted, undirected graph $\mathcal{G} = (V, E)$, where:
\begin{itemize}[itemsep=0pt]
    \item The set of vertices $V$ corresponds to the set of all $d$ channel indices, where each vertex represents a channel.
    \item Each vertex $i \in V$ has a node weight equal to its self-importance, calculated as $w_i = \| \mathbf{q}_i \mathbf{k}_i^T \|_F^2$.
    \item Each edge $(i, j) \in E$ has an edge weight equal to the interaction term between channels $i$ and $j$, calculated as $w_{ij} = 2 \cdot (\mathbf{q}_i^T \mathbf{q}_j) \cdot (\mathbf{k}_i^T \mathbf{k}_j)$.
\end{itemize}

With this construction, the total pruning error in \eqref{eq:decomposed_loss} is precisely the sum of all node and edge weights within the subgraph induced by the pruned vertex set $\mathcal{D}_{p}$. The channel pruning problem is thus transformed into finding a vertex subset $\mathcal{D}_{p}$ of size $n_{prune} = \lceil \lambda d \rceil$ that induces a subgraph with the minimum possible total weight.

Finding the exact solution to this minimum-weight induced subgraph problem is NP-hard. We therefore propose MIES (see Algorithm \ref{alg:iap_pruning}), an efficient greedy algorithm to find an effective approximate solution. The algorithm iteratively constructs the prune set $\mathcal{D}_{prune}$. At each step, it selects the channel that, when added to the set of already pruned channels, results in the smallest increase to the total cumulative error. The score $s_k$ for a candidate channel $k$ dynamically tracks this cumulative error, explicitly incorporating the weights of the edges connecting $k$ to all previously pruned channels. This ensures our selection process is guided by both channel self-importance (node weights) and inter-channel dependencies (edge weights), directly optimizing the objective in \eqref{eq:raw_opt}.

\begin{table*}[t!]
\centering
\scriptsize
\caption{Performance comparison on LLaMA-3-8B-Instruct at LongBench. GRACE consistently outperforms THINK across most LongBench subtasks under varying KV budgets. Best results for each setting are highlighted in bold.}
\label{tab:main}
\resizebox{\textwidth}{!}{
\begin{tabular}
{c@{\hspace{1.0em}} % <-- 新增：最左侧“Setting/KV”列
 l@{\hspace{0.05ex}}C{3.9em}@{\hspace{0.05ex}}C{3.9em}@{\hspace{0.05ex}}C{3.9em}@{\hspace{0.05ex}}
 c@{\hspace{0.00ex}}c@{\hspace{0.05ex}}C{3.9em}@{\hspace{0.05ex}}
 c@{\hspace{0.05ex}}C{3.9em}@{\hspace{0.05ex}}c@{\hspace{0.05ex}}C{3.9em}@{\hspace{0.05ex}}
 c@{\hspace{0.05ex}}c@{\hspace{0.05ex}}C{3.9em}@{\hspace{0.05ex}}
 c@{\hspace{0.15ex}}C{3.9em}@{\hspace{0.15ex}}C{3.9em}@{\hspace{0.35ex}}C{3.9em}}

\toprule
\multicolumn{2}{c}{\multirow{4}{*}{\textbf{Method}}} &
\multicolumn{3}{c}{\textbf{Single-Document QA}} &
\multicolumn{3}{c}{\textbf{Multi-Document QA}} &
\multicolumn{3}{c}{\textbf{Summarization}} &
\multicolumn{3}{c}{\textbf{Few-shot Learning}} &
\multicolumn{2}{c}{\textbf{Synthetic}} &
\multicolumn{2}{c}{\textbf{Code}} &
\multirow{4}{*}{\textbf{Avg.}} \\

\cmidrule(lr){3-5}\cmidrule(lr){6-8}\cmidrule(lr){9-11}\cmidrule(lr){12-14}\cmidrule(lr){15-16}\cmidrule(lr){17-18}
& &
\rotatebox[origin=c]{30}{\bf NrtvQA} &
\rotatebox[origin=c]{30}{\bf Qasper} &
\rotatebox[origin=c]{30}{\bf MF-en} &
\rotatebox[origin=c]{30}{\bf HotpotQA} &
\rotatebox[origin=c]{30}{\bf 2WikiMQA} &
\rotatebox[origin=c]{30}{\bf Musique} &
\rotatebox[origin=c]{30}{\bf GovReport} &
\rotatebox[origin=c]{30}{\bf QMSum} &
\rotatebox[origin=c]{30}{\bf MultiNews} &
\rotatebox[origin=c]{30}{\bf TREC} &
\rotatebox[origin=c]{30}{\bf TriviaQA} &
\rotatebox[origin=c]{30}{\bf SAMSum} &
\rotatebox[origin=c]{30}{\bf PRe~~} &
\rotatebox[origin=c]{30}{~\bf PCount} &
\rotatebox[origin=c]{30}{~\bf Lcc~~} &
\rotatebox[origin=c]{30}{~\bf RB-P~} \\

% =========================
% Group: KV-size 512 (共10行：5行SnapKV系 + 5行H2O系)
% =========================
\cmidrule{1-19}
\multirow{10}{*}{\rotatebox[origin=c]{90}{%
\parbox{2.4cm}{\centering\textbf{KV-size 512}}}} &
SnapKV & 24.85 & 23.65 & 37.20 & 43.05 & 34.30 & 20.30 & 22.63 & 22.71 & 23.89 & 69.50 & 90.39 & 40.38 & 69.50 & 5.81 & 60.36 & 56.22 & \cellcolor{lightgreen}40.30\\
& +THINK (0.5) & 24.61 & 25.54 & 36.97 & 41.51 & 32.39 & 20.83 & 21.37 & 22.70 & 23.79 & 69.00 & 90.31 & 39.85 & 69.50 & 5.84 & 61.37 & 59.22 & \cellcolor{lightgreen}40.30 \\
& +THINK (0.6)  & 24.84 & 23.49 & 37.48 & 40.42 & 33.15 & 19.43 & 20.66 & 22.10 & 22.73 & 59.00 & 90.37 & 37.31 & 69.50 & 6.39 & 59.62 & 58.72 & \cellcolor{lightgreen}39.07 \\
& +GRACE (0.5)  & 25.06 & 25.84 & 37.82 & 42.19 & 32.25 & 20.77 & 21.48 & 22.50 & 23.69 & 69.50 & 90.39 & 40.52 & 69.83 & 5.79 & 61.84 & 58.52 & \cellcolor{lightgreen}\textbf{40.50} \\
& +GRACE (0.6)  & 25.12 & 22.12 & 39.73 & 40.28 & 32.09 & 20.00 & 21.07 & 21.81 & 22.63 & 59.00 & 90.12 & 38.40 & 69.50 & 6.04 & 59.37 & 58.83 & \cellcolor{lightgreen}\textbf{39.13}\\
\cdashline{2-19}
& H2O           & 23.53 & 17.88 & 33.84 & 41.68 & 33.49 & 19.27 & 22.29 & 22.61 & 24.03 & 41.0  & 90.46 & 40.17 & 69.5  & 5.62 & 57.91 & 55.97 & \cellcolor{lightgreen}37.45\\
& +THINK (0.5)  & 24.21 & 16.39 & 34.96 & 39.47 & 30.41 & 19.17 & 21.58 & 22.31 & 22.93 & 41.00 & 90.16 & 40.55 & 69.08 & 5.37 & 59.69 & 58.36 & \cellcolor{lightgreen}37.23 \\
& +THINK (0.6)  & 23.52 & 14.62 & 34.99 & 38.41 & 30.86 & 19.60 & 20.76 & 22.06 & 21.65 & 40.00 & 89.14 & 38.44 & 69.50 & 5.36 & 58.09 & 57.25 & \cellcolor{lightgreen}\textbf{36.52} \\
& +GRACE (0.5)  & 24.34 & 16.41 & 35.12 & 39.57 & 30.14 & 19.13 & 21.26 & 22.40 & 22.92 & 41.00 & 90.08 & 40.64 & 69.20 & 5.52 & 59.79 & 58.42 & \cellcolor{lightgreen}\textbf{37.25} \\
& +GRACE (0.6)  & 23.92 & 15.58 & 34.20 & 39.85 & 30.37 & 19.32 & 20.83 & 22.37 & 21.72 & 39.51 & 88.85 & 38.56 & 69.01 & 5.34 & 57.56 & 57.25 & \cellcolor{lightgreen}\textbf{36.52} \\

\cmidrule{1-19}

% =========================
% Group: KV-size 2048 (同样10行)
% =========================
\multirow{10}{*}{\rotatebox[origin=c]{90}{%
\parbox{2.4cm}{\centering\textbf{KV-size 2048}}}} &
SnapKV & 25.86 & 29.59 & 39.52 & 43.84 & 36.05 & 21.36 & 25.90 & 23.38 & 26.24 & 73.50 & 90.56 & 41.65 & 69.25 & 5.17 & 60.04 & 55.86 & \cellcolor{lightgreen}41.74 \\
& +THINK (0.5) & 24.84 & 30.26 & 39.21 & 43.22 & 33.29 & 21.04 & 24.93 & 23.09 & 26.04 & 73.00 & 90.37 & 41.21 & 69.25 & 5.49 & 62.13 & 59.72 & \cellcolor{lightgreen}41.69 \\
& +THINK (0.6) & 24.64 & 28.57 & 40.54 & 41.17 & 31.07 & 21.51 & 23.32 & 23.04 & 24.94 & 72.00 & 90.36 & 38.67 & 69.50 & 6.07 & 59.32 & 59.28 & \cellcolor{lightgreen}40.87 \\
& +GRACE (0.5) & 24.87 & 30.27 & 39.47 & 43.59 & 33.22 & 21.51 & 25.29 & 22.92 & 26.15 & 73.50 & 90.37 & 40.59 & 69.08 & 5.54 & 61.93 & 59.89 & \cellcolor{lightgreen}\textbf{41.76} \\
& +GRACE (0.6) & 25.31 & 28.63 & 38.80 & 42.13 & 31.40 & 21.21 & 23.69 & 23.24 & 24.75 & 72.50 & 89.86 & 38.69 & 69.25 & 5.50 & 59.97 & 59.50 & \cellcolor{lightgreen}\textbf{40.90} \\
\cdashline{2-19}
& H2O          & 25.56 & 27.09 & 38.51 & 43.04 & 32.92 & 20.51 & 24.99 & 23.14 & 26.11 & 53.00 & 90.56 & 41.83 & 69.25 & 5.04 & 59.60 & 55.72 & \cellcolor{lightgreen}39.80 \\
& +THINK (0.5) & 25.01 & 25.59 & 38.79 & 42.27 & 31.26 & 20.46 & 23.71 & 23.34 & 25.64 & 53.00 & 90.37 & 41.29 & 69.50 & 5.20 & 61.70 & 59.16 & \cellcolor{lightgreen}39.77 \\
& +THINK (0.6) & 24.43 & 22.09 & 38.73 & 40.53 & 29.64 & 20.71 & 22.20 & 22.64 & 24.56 & 49.50 & 90.41 & 39.77 & 69.20 & 5.76 & 59.24 & 59.23 & \cellcolor{lightgreen}38.66 \\
& +GRACE (0.5) & 25.77 & 25.73 & 39.74 & 43.41 & 30.44 & 20.62 & 23.84 & 23.24 & 25.88 & 53.00 & 90.37 & 41.19 & 69.50 & 5.32 & 61.52 & 59.02 & \cellcolor{lightgreen}\textbf{39.91} \\
& +GRACE (0.6) & 24.42 & 22.99 & 39.55 & 41.23 & 30.37 & 20.65 & 22.17 & 22.55 & 24.47 & 49.00 & 89.61 & 39.12 & 69.50 & 5.83 & 60.06 & 58.25 & \cellcolor{lightgreen}\textbf{38.73} \\

\bottomrule
\end{tabular}}
\end{table*}

\subsection{Mitigating Query Discrepancy}
\label{sec:enhance}
\begin{figure}[!t]
  \centering
  \centerline{\includegraphics[width=0.9\linewidth]{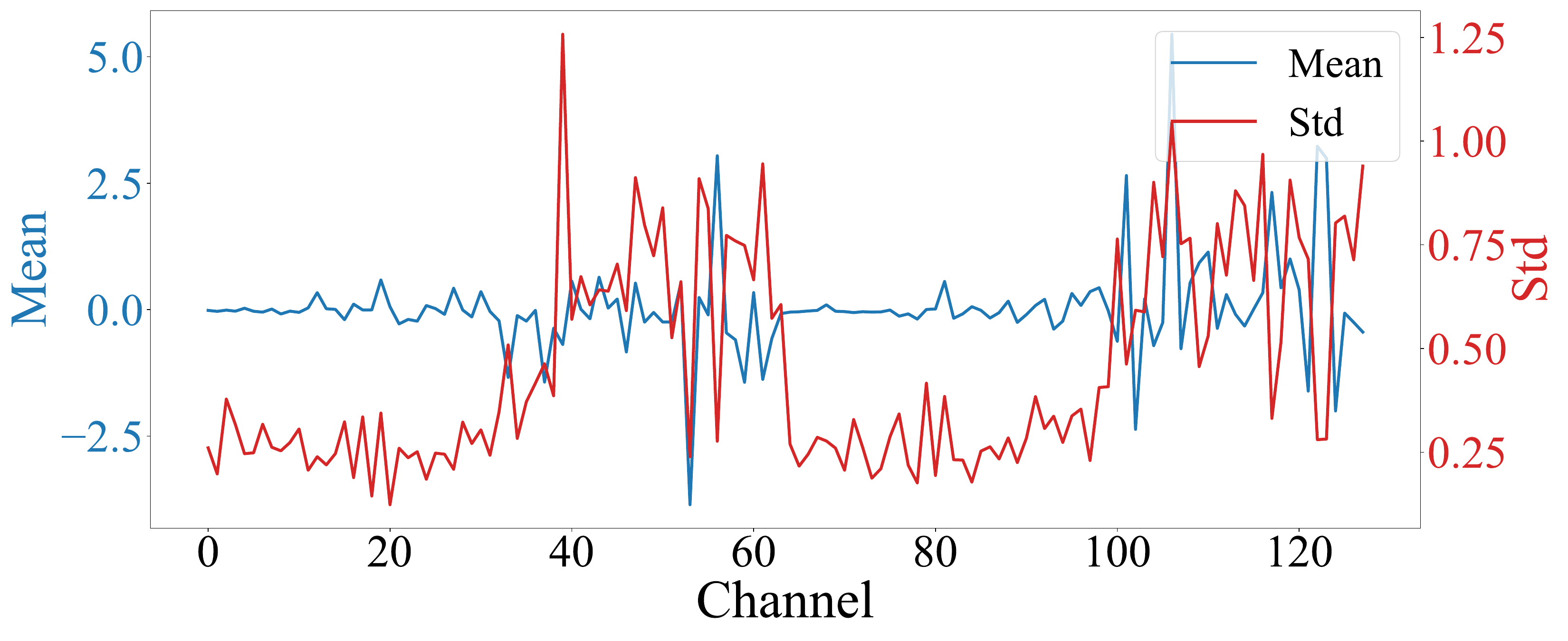}}
\caption{Statistical profiling of query vector channels. Channels with higher mean amplitudes also exhibit higher variance.}
\label{fig:mean_std}
\end{figure}
The optimization in \eqref{eq:raw_opt} is conditioned on an initial set of observed queries, $\mathbf{Q}_{0}^{obs}$. However, during autoregressive decoding, the cached keys $\mathbf{K}_{0}$ are probed by a sequence of evolving query vectors, $\mathbf{Q}_{t}$ ($t > 0$), creating a misalignment between the static pruning objective and the dynamic nature of inference. This discrepancy is non-trivial; as shown in Fig.~\ref{fig:mean_std}, query channels exhibit significant volatility, where channels with higher mean magnitudes also have greater variance. Consequently, future queries can diverge substantially from the initial sample, posing a significant risk of reconstruction error if high-magnitude key channels are prematurely pruned based on a limited initial observation window.

To mitigate the risk of distribution shift, and inspired by activation-aware quantization methods~\cite{lin2024awqactivationawareweightquantization}, we introduce a salient channel protection mechanism.
This mechanism preemptively excludes a subset of channels from the pruning candidate set. We identify these salient channels as those whose L2-norm exceeds a statistical threshold—specifically, the mean plus one standard deviation of all channel norms. However, the raw proportion of channels meeting this criterion, $p \in (0,1)$, can fluctuate with context. To ensure stable pruning, we clamp this proportion within adaptive bounds $[a, b]$: $p_{protect} = \text{min}(\text{max}(p,a),b)$.

These $p_{protect} \cdot d$ channels are then excluded from the candidate set $\mathcal{D}_{cand}$ in Algorithm~\ref{alg:iap_pruning}, guaranteeing their preservation.
This strategy serves two complementary purposes: it buffers against the volatility of future queries and explicitly safeguards ``Massive Activations''~\cite{sun2024massive, dettmers2022llmint88bitmatrixmultiplication}—outlier channels that act as structural pillars for model stability.

\subsection{Theoretical Analysis and Approximation Guarantees}
% As shown in \eqref{eq:decomposed_loss}, the reconstruction error for a pruned subset $S$ ($|S|=k$) is a quadratic set function $f(S) = \mathbf{1}_{S}^T W \mathbf{1}_{S}$, where $W$ is the interaction matrix.
% As a positive semi-definite Gram matrix, $W$ satisfies the restricted eigenvalue condition $\mu_{min} \|\mathbf{x}\|_2^2 \leq \mathbf{x}^T W \mathbf{x} \leq \mu_{max} \|\mathbf{x}\|_2^2$ for any $k$-sparse vector $\mathbf{x}$ (with $\mu_{max} \geq \mu_{min} > 0$). Consequently, our algorithm returns a subset $S_{MIES}$ satisfying:
% $$f(S_{MIES}) \leq \frac{\mu_{max}}{\mu_{min}} f(S^*)$$
% where $S^*$ is the exact optimal subset.

% This bound strictly justifies our salient channel protection. In LLMs, a few ``outlier'' channels possess extremely massive activations. If included in the candidate set, they would cause $\mu_{max}$ to explode, loosening the bound. By preemptively shielding them, GRACE flattens the eigenvalue spectrum, securely bounding the condition number $\kappa = \frac{\mu_{max}}{\mu_{min}}$ and theoretically guaranteeing a near-optimal subset $S_{MIES}$.

From \eqref{eq:decomposed_loss}, the reconstruction error for a pruned subset $S$ ($|S|=k$) can be written as a quadratic set function $f(S)=\mathbf{1}_S^T W \mathbf{1}_S$, where $W$ is a positive semi-definite interaction matrix.
Under the restricted eigenvalue condition
$\mu_{min}\|\mathbf{x}\|_2^2 \le \mathbf{x}^T W \mathbf{x} \le \mu_{max}\|\mathbf{x}\|_2^2$
for any $k$-sparse vector $\mathbf{x}$, our method returns a subset $S_{MIES}$ satisfying
$$
f(S_{MIES}) \le \frac{\mu_{max}}{\mu_{min}} f(S^*),
$$
where $S^*$ denotes the optimal subset.

This result explains salient channel protection: outlier channels in LLMs enlarge $\mu_{max}$ and weaken the bound. By shielding them beforehand, GRACE stabilizes the eigenvalue spectrum, controls the condition number $\kappa=\mu_{max}/\mu_{min}$, and guarantees a near-optimal solution.

\begin{figure}[!t]
  \centering
  \centerline{\includegraphics[width=0.9\linewidth]{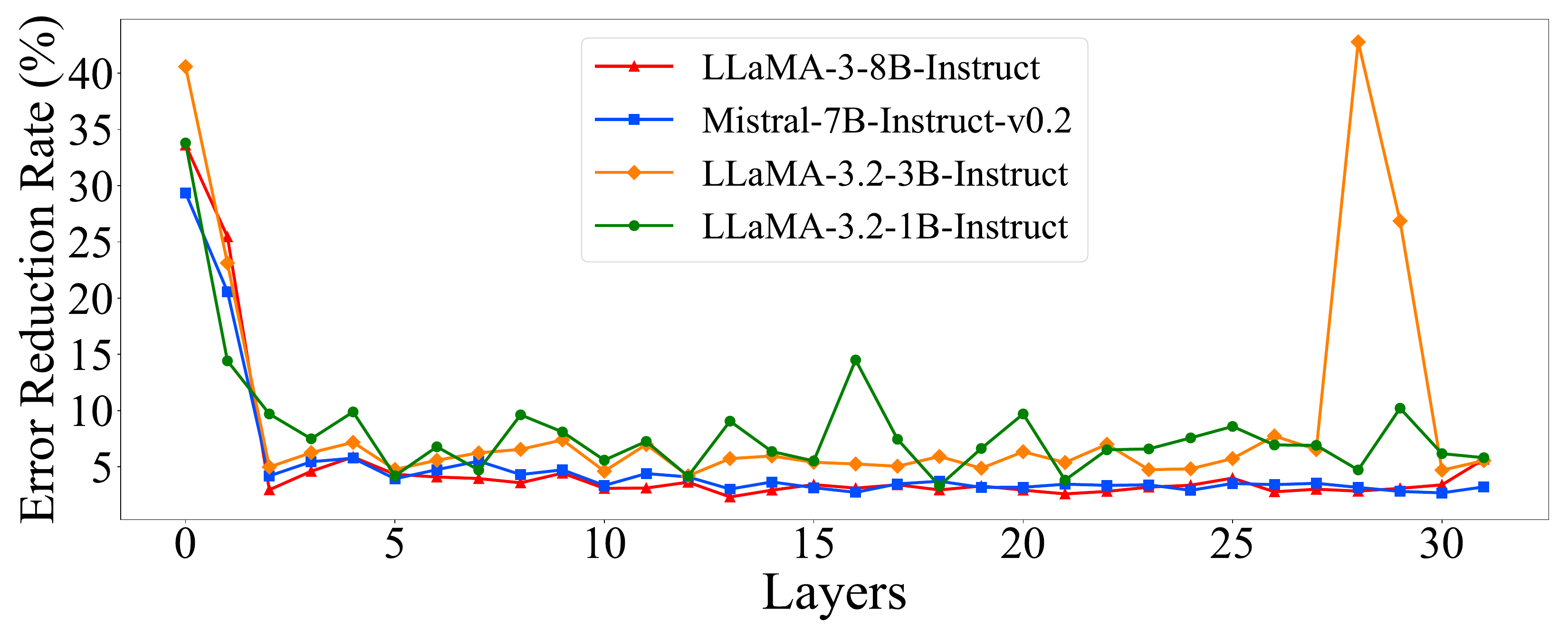}}
\caption{Reduction in reconstruction error for our method versus the THINK across decoder layers.}
\label{fig:diff}
\end{figure}

\section{Experiments}

\begin{table*}[t!]
\centering
\scriptsize
\caption{Performance comparison on Mistral-7B-Instruct at LongBench. GRACE consistently outperforms THINK across most LongBench subtasks under varying KV budgets. Best results for each setting are highlighted in bold.}
\label{tab:main_mistral}
\resizebox{\textwidth}{!}{
\begin{tabular}
{c@{\hspace{1.0em}}
 l@{\hspace{0.05ex}}C{3.9em}@{\hspace{0.05ex}}C{3.9em}@{\hspace{0.05ex}}C{3.9em}@{\hspace{0.05ex}}
 c@{\hspace{0.00ex}}c@{\hspace{0.05ex}}C{3.9em}@{\hspace{0.05ex}}
 c@{\hspace{0.05ex}}C{3.9em}@{\hspace{0.05ex}}c@{\hspace{0.05ex}}C{3.9em}@{\hspace{0.05ex}}
 c@{\hspace{0.05ex}}c@{\hspace{0.05ex}}C{3.9em}@{\hspace{0.05ex}}
 c@{\hspace{0.15ex}}C{3.9em}@{\hspace{0.15ex}}C{3.9em}@{\hspace{0.35ex}}C{3.9em}}

\toprule
\multicolumn{2}{c}{\multirow{4}{*}{\textbf{Method}}} &
\multicolumn{3}{c}{\textbf{Single-Document QA}} &
\multicolumn{3}{c}{\textbf{Multi-Document QA}} &
\multicolumn{3}{c}{\textbf{Summarization}} &
\multicolumn{3}{c}{\textbf{Few-shot Learning}} &
\multicolumn{2}{c}{\textbf{Synthetic}} &
\multicolumn{2}{c}{\textbf{Code}} &
\multirow{4}{*}{\textbf{Avg.}} \\

\cmidrule(lr){3-5}\cmidrule(lr){6-8}\cmidrule(lr){9-11}\cmidrule(lr){12-14}\cmidrule(lr){15-16}\cmidrule(lr){17-18}
& &
\rotatebox[origin=c]{30}{\bf NrtvQA} &
\rotatebox[origin=c]{30}{\bf Qasper} &
\rotatebox[origin=c]{30}{\bf MF-en} &
\rotatebox[origin=c]{30}{\bf HotpotQA} &
\rotatebox[origin=c]{30}{\bf 2WikiMQA} &
\rotatebox[origin=c]{30}{\bf Musique} &
\rotatebox[origin=c]{30}{\bf GovReport} &
\rotatebox[origin=c]{30}{\bf QMSum} &
\rotatebox[origin=c]{30}{\bf MultiNews} &
\rotatebox[origin=c]{30}{\bf TREC} &
\rotatebox[origin=c]{30}{\bf TriviaQA} &
\rotatebox[origin=c]{30}{\bf SAMSum} &
\rotatebox[origin=c]{30}{\bf PRe~~} &
\rotatebox[origin=c]{30}{~\bf PCount} &
\rotatebox[origin=c]{30}{~\bf Lcc~~} &
\rotatebox[origin=c]{30}{~\bf RB-P~} \\

\cmidrule{1-19}
\multirow[c]{5}{*}{\rotatebox[origin=c]{90}{%
\shortstack[c]{\textbf{KV-size 256}}}}  &
    SnapKV & 22.34 & 23.71 & 48.39 & 38.73 & 22.57 & 15.56 & 21.59 & 22.96 & 23.00 & 61.5 & 85.68 & 41.26 & 85.54 & 3.04 & 54.95 & 51.90 & \cellcolor{lightgreen}38.92 \\
    & +THINK (0.5) & 21.92 & 24.74 & 47.58 & 38.48 & 22.10 & 15.88 & 21.47 & 22.92 & 22.51 & 58.50 & 85.92 & 40.98 & 84.48 & 2.87 & 54.53 & 50.80 & \cellcolor{lightgreen}38.48 \\
    & +THINK (0.6) & 22.29 & 24.75 & 47.44 & 38.18 & 22.19 & 15.11 & 21.32 & 23.31 & 22.83 & 55.50 & 85.16 & 40.20 & 82.83 & 3.21 & 53.47 & 50.27 & \cellcolor{lightgreen}38.00 \\
    & +GRACE (0.5) & 22.18 & 24.15 & 47.53 & 38.12 & 21.43 & 16.85 & 21.55 & 23.07 & 22.75 & 60.00 & 85.99 & 40.38 & 83.85 & 3.02 & 54.28 & 51.24 & \cellcolor{lightgreen}\textbf{38.52} \\
    & +GRACE (0.6) & 23.07 & 24.10 & 47.82 & 38.63 & 21.69 & 16.02 & 21.13 & 22.91 & 22.58 & 58.00 & 85.25 & 40.15 & 82.48 & 2.78 & 53.38 & 50.00 & \cellcolor{lightgreen}\textbf{38.12} \\
\cmidrule{1-19}
\multirow[c]{5}{*}{\rotatebox[origin=c]{90}{%
\shortstack[c]{\textbf{KV-size 512}}}} &
    SnapKV & 24.23 & 27.94 & 48.76 & 39.56 & 25.09 & 17.42 & 23.41 & 23.13 & 24.66 & 67.00 & 85.88 & 41.47 & 86.38 & 2.87 & 56.45 & 53.35 & \cellcolor{lightgreen}40.48 \\
    & +THINK (0.5) & 24.29 & 28.43 & 49.03 & 38.54 & 24.77 & 16.55 & 23.52 & 23.47 & 24.35 & 66.50 & 85.83 & 42.28 & 86.07 & 3.21 & 56.16 & 52.83 & \cellcolor{lightgreen}40.36 \\
    & +THINK (0.6) & 23.92 & 28.41 & 49.17 & 38.78 & 24.36 & 16.45 & 23.16 & 23.07 & 24.04 & 65.50 & 85.48 & 41.17 & 84.65 & 3.20 & 54.87 & 52.25 & \cellcolor{lightgreen}39.90 \\
    & +GRACE (0.5) & 23.85 & 28.80 & 48.90 & 38.65 & 24.30 & 16.69 & 23.48 & 23.72 & 24.30 & 67.00 & 85.88 & 41.65 & 85.95 & 3.10 & 56.35 & 53.31 & \cellcolor{lightgreen}\textbf{40.37} \\
    & +GRACE (0.6) & 23.88 & 28.58 & 49.33 & 38.41 & 23.78 & 16.98 & 22.64 & 23.20 & 24.15 & 66.50 & 85.82 & 40.85 & 85.18 & 2.84 & 55.32 & 52.13 & \cellcolor{lightgreen}\textbf{39.97} \\
\bottomrule
\end{tabular}}
\end{table*}

\subsection{Experimental Setup}
Our experiments are conducted on two leading open source models: LLaMA-3-8B-Instruct and Mistral-7B-Instruct-v0.2, obtained from the HuggingFace Hub \cite{wolf2020huggingfacestransformersstateoftheartnatural}. We evaluate performance on the LongBench \cite{bai2024longbenchbilingualmultitaskbenchmark} and Needle In A Haystack \cite{LLMTest_NeedleInAHaystack}. LongBench is a comprehensive benchmark consisting of six diverse task categories for evaluating long-context understanding in LLMs. The Needle In A Haystack test measures a model's ability to recall a specific fact from a long and noisy context, directly assessing its long-context retrieval capabilities.

\subsection{Comprehensive Experimental Analysis}
\begin{table}[!t]
\centering
\small
\caption{Average retrieval scores on the Needle In A Haystack for LLaMA-3-8B. Best results are highlighted in bold.}
\label{tab:needle_results}
\renewcommand{\arraystretch}{0.5}
\begin{tabular*}{\columnwidth}{c @{\extracolsep{\fill}} ccc}
\toprule
\textbf{KV Cache Size} & \textbf{Pruning Ratio} & \textbf{Method} & \textbf{Avg. Score} \\
\midrule
\multirow{4}{*}{96} & \multirow{2}{*}{0.5} & THINK & 0.873 \\
                    &                      & GRACE   & \textbf{0.876} \\
\cmidrule(lr){2-4}
                    & \multirow{2}{*}{0.6} & THINK & 0.804 \\
                    &                      & GRACE   & \textbf{0.828} \\
\midrule
\multirow{4}{*}{256} & \multirow{2}{*}{0.5} & THINK & 0.889 \\
                     &                      & GRACE   & \textbf{0.891} \\
\cmidrule(lr){2-4}
                     & \multirow{2}{*}{0.6} & THINK & 0.810 \\
                     &                      & GRACE   & \textbf{0.818} \\
\midrule
\multirow{4}{*}{512} & \multirow{2}{*}{0.5} & THINK & \textbf{0.975} \\
                     &                      & GRACE   & \textbf{0.975} \\
\cmidrule(lr){2-4}
                     & \multirow{2}{*}{0.6} & THINK & 0.918 \\
                     &                      & GRACE   & \textbf{0.920} \\
\bottomrule
\end{tabular*}
\end{table}
\noindent\textbf{Analysis of Attention Weight Reconstruction Error:}
To validate GRACE's optimization fidelity, we evaluated its ability to minimize the attention reconstruction error defined in \eqref{eq:raw_opt}. We compared GRACE against THINK on the Qasper dataset, with results shown in Fig.~\ref{fig:diff}. GRACE consistently achieves a lower reconstruction error, particularly in the shallowest and deepest layers where it secures a 30-40\% relative reduction. This confirms that GRACE's graph-based approach more effectively identifies an optimal channel subset by modeling inter-channel interactions.

\noindent\textbf{Performance on LongBench:}
To evaluate the efficacy of our method on diverse long-context tasks, we benchmarked GRACE against THINK on LongBench using both LLaMA-3-8B-Instruct (Table~\ref{tab:main}) and Mistral-7B-Instruct (Table~\ref{tab:main_mistral}).
On LLaMA-3-8B, GRACE consistently surpasses THINK across various KV cache sizes and pruning ratios. For instance, with a 512-token KV budget (using SnapKV) and a 50\% pruning ratio, GRACE achieves a superior average score of 40.50 compared to THINK's 40.30, outperforming it on 10 out of 16 subtasks.
This robustness extends to the Mistral-7B model. As shown in Table~\ref{tab:main_mistral}, GRACE (0.6) maintains a clear advantage over THINK (0.6), delivering higher average scores under both 256-token (38.12 vs. 38.00) and 512-token (39.97 vs. 39.90) budgets.
Collectively, these results validate that GRACE's graph-theoretic approach generalizes well across different model architectures and compression constraints, effectively preserving critical information where independent scoring methods falter.

\begin{table}[!t]
    \centering
    \small
    \caption{Inference efficiency comparison between GRACE and THINK on LLaMA-3-8B, measured by Time To First Token (TTFT) and Time Per Output Token (TPOT).}
    \label{tab:ttft}
    \begin{tabular*}{\linewidth}{@{\extracolsep{\fill}}lcccc}
        \toprule
        \textbf{Metrics} & \makecell{\textbf{THINK}\\\textbf{(0.5)}} & \makecell{\textbf{GRACE}\\\textbf{(0.5)}} & \makecell{\textbf{THINK}\\\textbf{(0.6)}} & \makecell{\textbf{GRACE}\\\textbf{(0.6)}} \\
        \midrule
        TTFT (s) & 2.42 & 2.73 & 2.33 & 2.80 \\
        TPOT (ms/token) & 72.24 & 71.66 & 72.09 & 72.37 \\
        \bottomrule
    \end{tabular*}
\end{table}

% \vspace{\baselineskip}

\noindent\textbf{Performance on Needle In A Haystack:}
To measure factual recall under aggressive compression, we employed the Needle In A Haystack test, which challenges the model to retrieve specific information from a long context (detailed in Table~\ref{tab:needle_results}). GRACE consistently outperforms THINK, with its advantage being most pronounced in highly constrained settings. With a minimal 96-token KV cache and a 60\% pruning ratio, GRACE achieves a retrieval score of 0.828, significantly surpassing THINK's 0.804. This demonstrates that GRACE excels at preserving critical, sparse information, making it highly effective for extreme compression scenarios.

% \vspace{\baselineskip}

\noindent\textbf{Inference Efficiency:}
To quantify the computational overhead, we benchmarked the inference efficiency of GRACE against THINK on a single NVIDIA RTX 3090 GPU (detailed in Table~\ref{tab:ttft}). The results show that GRACE introduces only a marginal increase in the Time to First Token (TTFT) while maintaining a highly comparable Time per Output Token (TPOT), which measures sustained generation speed. This key finding indicates that the significant performance improvements afforded by GRACE are achieved with virtually negligible impact on overall inference throughput.

\section{Conclusion}
In this work, we introduced a novel channel-level KV cache pruning method that outperforms existing approaches on long-context benchmarks. Our method's effectiveness stems from its unique graph-based formulation, which explicitly models the inter-channel interactions that are critical to performance, and a protection mechanism for salient channels that ensures robustness against query distribution shifts. Extensive experiments on LongBench and Needle In A Haystack demonstrate the effectiveness of our method.

\bibliographystyle{IEEEbib}
\bibliography{icme2026references}

\end{document}